\documentclass[10pt,twocolumn,letterpaper]{article}

\usepackage[pagenumbers]{cvpr} 
\usepackage{media9}
\usepackage{animate}
\usepackage[accsupp]{axessibility}
\usepackage{algorithm}
\usepackage{algpseudocode}

\usepackage[dvipsnames]{xcolor}

\definecolor{cvprblue}{rgb}{0.21,0.49,0.74}
\usepackage[pagebackref,breaklinks,colorlinks,citecolor=cvprblue]{hyperref}

\def\paperID{3747} 
\def\confName{CVPR}
\def\confYear{2025}

\title{AnyMoLe: Any Character Motion In-betweening \\ Leveraging Video Diffusion Models}

\author{
Kwan Yun \quad  Seokhyeon Hong \quad Chaelin Kim \quad  Junyong Noh
 \vspace{1mm}
 \\KAIST, Visual Media Lab
 \\{\tt\small\{ yunandy, ghd3079, chaelin.kim, junyongnoh\}}\tt\small@kaist.ac.kr
}
\usepackage{color, xcolor, soul}

\begin{document}
\maketitle

\begin{abstract}
Despite recent advancements in learning-based motion in-betweening, a key limitation has been overlooked: the requirement for character-specific datasets. In this work, we introduce AnyMoLe, a novel method that addresses this limitation by leveraging video diffusion models to generate motion in-between frames for arbitrary characters without external data. Our approach employs a two-stage frame generation process to enhance contextual understanding. Furthermore, to bridge the domain gap between real-world and rendered character animations, we introduce ICAdapt, a fine-tuning technique for video diffusion models. Additionally, we propose a ``motion-video mimicking'' optimization technique, enabling seamless motion generation for characters with arbitrary joint structures using 2D and 3D-aware features. AnyMoLe significantly reduces data dependency while generating smooth and realistic transitions, making it applicable to a wide range of motion in-betweening tasks. The code and videos are available at {\href{https://kwanyun.github.io/AnyMoLe_page/}{project page}}.
\end{abstract}    
\vspace{-5mm}
\section{Introduction}
\label{sec:intro}
Keyframe interpolation is essential in the animation creation process, providing smooth and natural transitions between keyframes to achieve lifelike character movements. Traditionally, this process has been highly labor-intensive, often involving extensive manual adjustments to refine the animation. In recent years, while the adoption of motion capture (MOCAP) technologies has helped alleviate this manual burden, its high costs and infeasibility for certain characters and animals present significant challenges. To address these limitations, deep learning-based motion in-betweening methods have emerged, making notable advancements~\cite{harvey2020robust, duan2021single}. These efforts have focused on developing sophisticated architectures~\cite{oreshkin2023motion, qin2022motion, duan2022unified, cohan2024flexible} and utilizing large datasets~\cite{harvey2020robust} to generate realistic motion sequences over long durations. Nevertheless, a major hurdle persists: training existing in-betweening models still demands vast amounts of MOCAP or manually keyframed data for each character. This is a crucial issue, as motion in-betweening is not only valuable for well-documented characters but is even more essential for those that are difficult to capture with MOCAP or lack extensive keyframed data.

\begin{figure}[t]
    \centering
    \hspace{-2mm}
\begin{minipage}[t]{0.49\linewidth}
    \animategraphics[autoplay,loop,width=1\linewidth,trim=80 0 50 0,clip]{15}{media/raptor_jpg/}{1}{59}
\end{minipage}
\begin{minipage}[t]{0.51\linewidth}
    \animategraphics[autoplay,loop,width=1\linewidth,trim=40 0 90 0,clip]{15}{media/mutant_jpg/}{1}{59}
\end{minipage} \vspace{-7mm}
    \caption{AnyMoLe generates in-between motion from context frames and keyframes without requiring external training data.}\label{fig:teaser_vis}
    \vspace{-6mm}
\end{figure}

Recent advancements in video generation models, trained on web-scale video datasets, have enabled the generation of realistic videos from arbitrary text~\cite{blattmann2023stable,blattmann2023align,guo2023animatediff,videoworldsimulators2024} or images~\cite{xing2023dynamicrafter,zhang2023i2vgen}. Additionally, some models have demonstrated the ability to interpolate between two images, naturally generating a range of transitioning scene elements or camera movements~\cite{xing2023dynamicrafter, xing2024tooncrafter, danier2024ldmvfi}. These capabilities align closely with the goal of motion in-betweening, as video generation models excel at creating smooth transitions and continuous motion. Therefore, these models can address data scarcity issues often encountered in 3D motion in-betweening and leverage the generation of realistic motions from rendered keyframes, which become the essential foundation of our motion in-betweening process.


We propose a novel motion in-betweening method, AnyMoLe, which addresses the problem of scarcity of character-specific datasets. AnyMoLe leverages video diffusion models to generate in-betweening motion through a sequential process of rendering, video generation, and motion optimization. In this process, we observed that naively generating an interpolated video and reconstructing a 3D character often yielded unsatisfactory results due to: 1) limited contextual understanding, 2) domain gaps between real-world and rendered scenes, and 3) difficulties in accurately tracking motion from a generated video. To address these issues, we propose components specifically designed to overcome each challenge.

Limited contextual understanding arises from the fixed input of video diffusion models, which only utilize the first and last frames, without direct knowledge of the intermediate or previous frames. To address this lack of contextual awareness, we condition frame generation on previous frames as motion context, treating the problem as frame inpainting by incorporating masked noise during the video generation process. Additionally, we employ a two-stage approach for smooth video generation: first, generating sparse frames to establish the motion structure, followed by dense frame generation to fill in the details.

The domain gap stems from differences in the distributions between real-world videos and rendered scenes. Video diffusion models are typically trained on real-world videos, which often feature elements like motion blur and complex organic environments. In contrast, rendered scenes usually depict synthetic environments and virtual characters. To bridge this gap, we employ Inference-stage Context Adaptation (ICAdapt). This approach fine-tunes the video diffusion model by using only a short, 2-second segment of context motion. During this process, the spatial module is overfitted to accurately represent specific virtual characters, while the temporal module remains frozen to preserve the learned motion dynamics.

To address the difficulty in tracking the motion of arbitrary rigged characters in the generated video, we propose a novel motion-video mimicking method. Motion-video mimicking is similar to 3D reconstruction with a difference in allowed degree of freedom, as it is for rigged characters with an arbitrary joint structure. For motion-video mimicking, we propose a sequential optimization approach that predicts the states of 3D joints to generate smooth transitions. The joint prediction is performed by our newly proposed scene-specific joint estimator, specifically designed for motion in-betweening in a few-shot setting, using only the context frames and keyframes for training.

\vspace{-1.5mm}
\begin{table}[h]
    \centering
    \caption{Difference between AnyMoLe and baseline methods.}
    \vspace{-2.5mm}
    \label{tab:diff_baseline}
    \scalebox{0.8}{
        \begin{tabular}{lcccc}
            \hline
            \textbf{Methods} &  Characters & Training data & Output  \\ \hline
            AnyMoLe                      & Arbitrary & None & 3D motion \\ \hline
            TS~\cite{qin2022motion} & Human & 3D motion Dataset & 3D motion \\ \hline
            Deciwatch~\cite{zeng2022deciwatch}    & Human & video-3D poses Dataset & 3D poses \\ \hline
        \end{tabular}
    }\vspace{-2mm}
\end{table}

By integrating these techniques, AnyMoLe successfully performs the first motion in-betweening for arbitrary characters as shown in Figure~\ref{fig:teaser_vis}. The differences between AnyMoLe and previous methods are highlighted in Table~\ref{tab:diff_baseline}. Our contributions can be summarized as follows:
\begin{itemize}
    \item By utilizing a video diffusion model with two stage inference for contextual understanding, we propose the first motion in-betweening method for arbitrary characters without external data.
    \item We present ICAdapt, to bridge the domain gaps between real-world videos and rendered scenes by finetuning a video diffusion model.
    \item We introduce a novel optimization technique for motion-video mimicking, which enables sequential motion optimization for arbitrary characters.
    \item We propose a new scene-specific joint estimator, specialized for a specific rendering scene, which utilizes 2D and 3D-aware features for effective 3D joint estimation.
\end{itemize}



\vspace{-2mm}
\section{Related Work}\vspace{-1mm}
\label{sec:related}
\vspace{-1mm}
\subsection{Motion In-betweening}
Motion in-betweening aims to generate smooth and natural motion transitions by a 3D character conditioned on a sparse set of keyframes.
Early approaches mainly utilized optimization with space-time constraints~\cite{rose1996efficient, witkin1988spacetime} or radial basis functions~\cite{rose1998verbs, rose2001artist} to interpolate keyframes.
Other approaches leveraged motion graphs~\cite{kovar2002motion, lee2002interactive} constructed from a given motion dataset, where optimal paths connecting two keyframes are found within these graphs~\cite{kovar2004automated}.
Statistical models such as maximum a posteriori~\cite{chai2007constraint} and geostatistical models~\cite{mukai2005geostatistical} were also employed for transition synthesis.

With the rise of learning-based methods, neural network models have produced promising results for motion in-betweening.
Specifically, Recurrent Neural Networks~\cite{harvey2018recurrent, harvey2020robust, kim2023recurrent, tang2022real, tang2023rsmt}, Convolutional Neural Networks~\cite{kaufmann2020convolutional, hernandez2019human, zhou2020generative, li2021task}, and Transformers~\cite{qin2022motion, oreshkin2023motion, hong2024long} have been widely adopted.
Recently, diffusion-based models have gained popularity for their ability to generate intermediate transitions by leveraging their generative capabilities~\cite{tevet2023human, kim2022flame, cohan2024flexible}.
Despite these advancements, performing motion in-betweening for diverse characters within a single model remains challenging, because preparing extensive animation data for each specific character is infeasible.
In contrast, our method harnesses the capabilities of video diffusion models to create in-between frames for any characters with arbitrary joint structures, bypassing the need for preparing extensive motion data.

\vspace{-2mm}
\subsection{Diffusion Based Video Interpolation Models}
\vspace{-1mm}

Recent advancements in image-conditioned video diffusion models enable the generation of promising results for video frame interpolation, particularly addressing the limitations of traditional methods. Diffusion models, originally applied to large-scale text-to-video (T2V) generation~\cite{singer2022make,xing2024make,ho2022imagenvideo,wang2024recipe,blattmann2023align}, have been extended to image-to-video (I2V) synthesis~\cite{gu2023seer,wang2024videocomposer,zhang2023i2vgen,xing2023dynamicrafter,blattmann2023stable}. Techniques such as SEINE~\cite{chen2023seine} and PixelDance~\cite{zeng2023makepixeldance} leverage these models to generate image transitions, while DynamiCrafter~\cite{xing2023dynamicrafter} utilize them for frame interpolation between two input images. Most recently, ToonCrafter~\cite{xing2024tooncrafter} finetuned a general interpolation model, to become cartoon interpolation model using a relatively small dataset of cartoon videos (a few hundred thousand samples compared to the original dataset of ten millions). Similarly, we finetune a video diffusion model but with only two seconds of context motion for single motion adaptation.

\begin{figure*}[t]
\centering
\vspace{-5mm}
\includegraphics[width=1\linewidth]{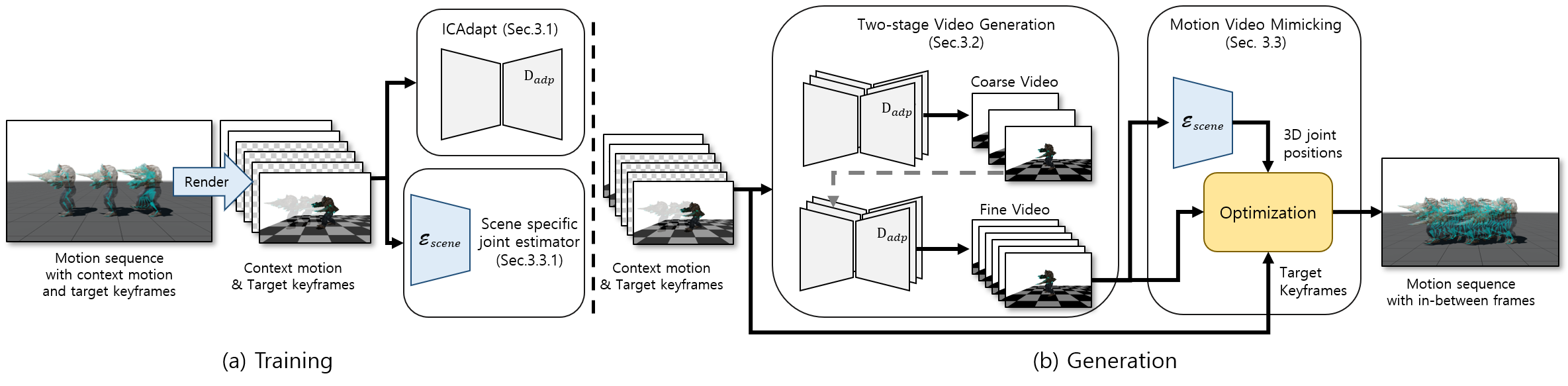}
\vspace{-5mm}
\caption{
Overview of AnyMoLe: First, the video diffusion model is fine-tuned without using any external data (Sec.~\ref{subsec:ICAdapt}) while the scene-specific joint estimator is trained (Sec.~\ref{subsec:pose}). Next, the fine-tuned video generation model produces an in-between video (Sec.~\ref{subsec:videogen}), which is then refined through motion video mimicking to generate the final in-between motion (Sec.~\ref{subsec:mimicking}).
}
\vspace{-2mm}
\label{fig:overview}
\end{figure*}

\section{Methods}
Given two seconds of context motion and target key frames for a 3D character, AnyMoLe performs 3D motion in-betweening using a video diffusion model, as illustrated in Figure~\ref{fig:overview}. The process can be described as follows:

\begin{enumerate}  
\item Using the two seconds of context motion and target keyframes, render each frame from diverse views.

\item From the multi-view rendered images, finetune the video diffusion model and train a scene-specific joint estimator concurrently using both 2D and 3D-aware features.

\item The finetuned video diffusion model generates an in-between video in a two stage auto-regressive manner: first, generate coarse frames from the given context and key frames, second, fill in the remaining frames.

\item Optimize character motion sequentially to align it motion with the generated video using a differentiable renderer and the trained joint estimator.

\end{enumerate}
In the following subsections, we will describe the ICAdapt (Sec.~\ref{subsec:ICAdapt}), two stage conditional video generation process (Sec.~\ref{subsec:videogen}), and progressive optimization process for motion-video mimicking (Sec.~\ref{subsec:mimicking}).

\subsection{Inference-stage Context Adaptation}\label{subsec:ICAdapt}
We use DynamiCrafter~\cite{xing2023dynamicrafter}, a state-of-the art open-source video interpolation model as our baseline. While this video diffusion model has demonstrated robust motion understanding capability for live-action videos, there is a domain gap when applying it to rendered character animation.
For example, rendered scenes contain virtual avatars and synthetic backgrounds that differ from the scenes observed in real-world videos, and they lack camera effects such as motion blur. Therefore, we finetune the video diffusion model at the start of inference, using the videos rendered from two seconds of given context motions. If we train the whole video diffusion model using a single motion, it will lead to catastrophic forgetting which was evident in~\citet{xing2024tooncrafter}. Therefore, we finetune the spatial module and image projector of DynamiCrafter while preserving the original temporal module and fps embedding. This process ensures that finetuned model $D_{adp}$ faithfully generates a video as it is rendered while keeping its ability to generate new motions.

\begin{figure}[h]
\centering
\hspace{-6mm}
\includegraphics[width=1.05\linewidth]{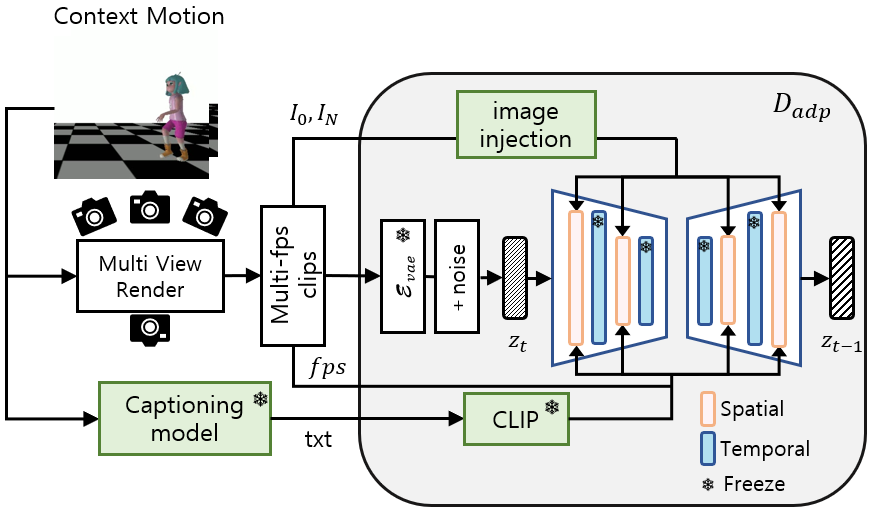}
\vspace{-2mm}
\caption{Overview of the ICAdapt training process. The spatial module and image injection module are trained, while the others are frozen.}
\label{fig:icadapt}
\end{figure}

Specifically, for the input, we render two seconds of motion at 30 fps from $N$ different views, resulting in $N$ videos. Then, we sample images with intervals of 1, 2, and 3 frames, which becomes 30, 15, and 10 fps videos, respectively. We segment each video into 16 frame intervals using a sliding window to align with the output dimension of $D_{adp}$. As a result, the two seconds motion is converted to multiple video clips, each containing 16 frames, aligning with the output dimension of the diffusion model. Finetuning the spatial module and image projector with these clips ensures that the generated videos fall in the domain of rendering scene while preserving the details of the input character. 

Figure~\ref{fig:icadapt} shows the process for ICAdapt, whose objective function is defined as follows:
\begin{equation}
\min_{\theta} \mathbb{E}_{\mathcal{E}_{vae}(\mathbf{x}), t, \mathbf{\epsilon}} \left[\Vert\mathbf{\epsilon}-\mathbf{\epsilon}_{\theta}\left(\mathbf{z}_t;{I_0, I_N},\mathbf{txt},t,\textit{fps}\right)\Vert_2^2 \right],
\end{equation}
where $ \mathbf{\epsilon} \sim \mathcal{N}(\mathbf{0}, \mathbf{I})$ represents noise, $I_0$ and $I_N$ denote the first and last frames, $\mathbf{txt}$ is a text condition automatically generated from a pretrained video captioning model~\cite{yu2024eliciting}, and $\textit{fps}$ refer to the frame rate. This training aims to capture the spatial information that remains constant (e.g., scene environment and character) while avoiding overfitting to the temporal information that may change (e.g., walking to running). The training is conducted in latent space, as $D_{adp}$ is a latent diffusion model.

\subsection{Two-stage Video Generation}\label{subsec:videogen}
\begin{figure}[t]
\centering
\vspace{-3mm}
\includegraphics[width=0.95\linewidth]{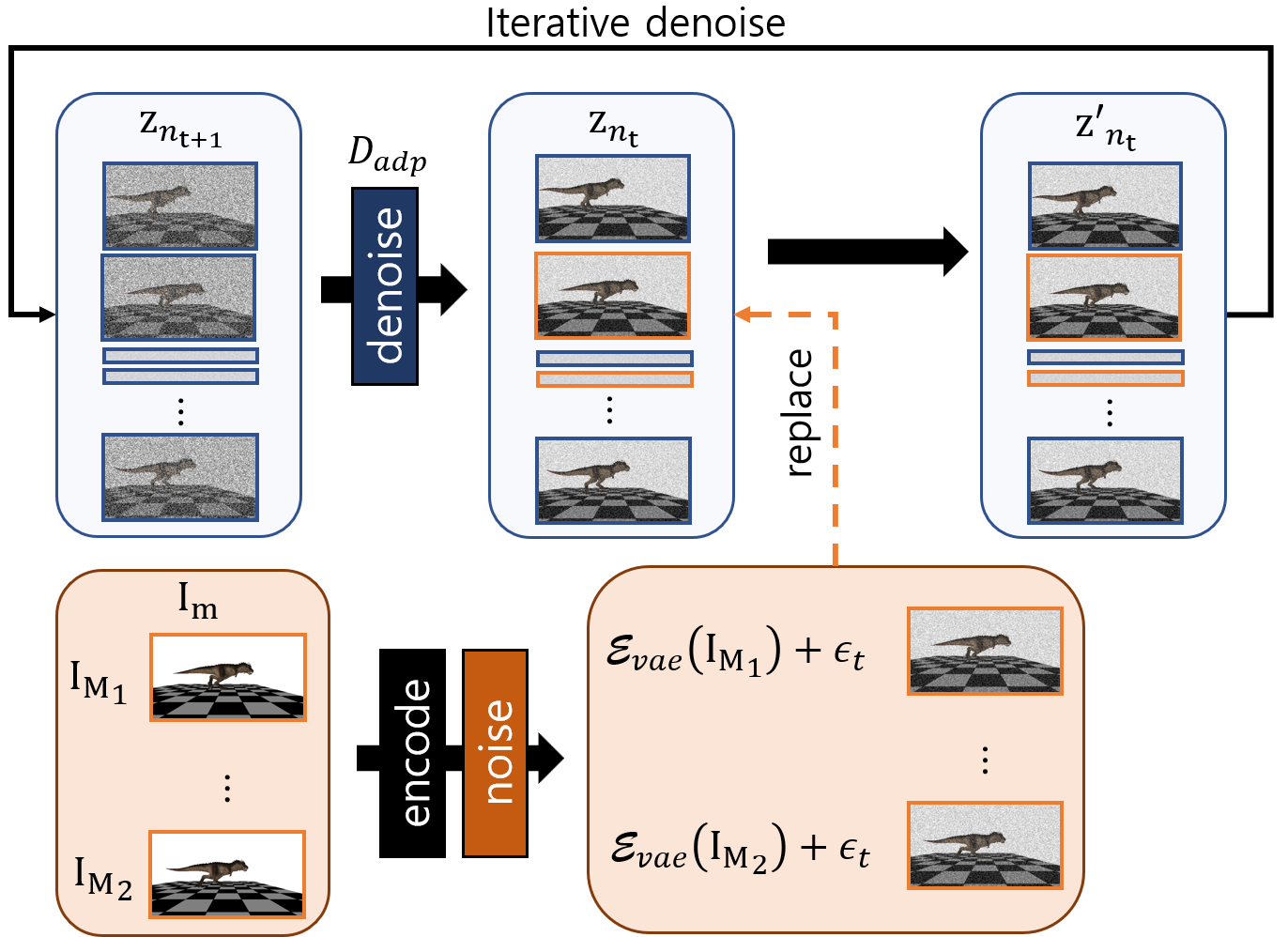}
\caption{Context frames guided video generation process.}
\vspace{-3mm}
\label{fig:inpaint}
\end{figure}

With the finetuned video diffusion model $D_{adp}$, we can now generate an interpolated video for the target character, a recipe for our final in-betweened motion. Here, naively generating a video from keyframes would produce results with wrong context, as interpolation lacks contextual information about the previous motion. To generate a video that naturally reflects the given context, we first produce coarse frames by using context frames as guidance in an auto-regressive manner. Because $D_{adp}$ is a video interpolation model that originally takes only the first and last images as input and cannot utilize additional context frames, we reformulate the task of interpolation with context frame guidance as a problem of latent inpainting during the diffusion process~\cite{lugmayr2022repaint,avrahami2022blended}. Specifically, $D_{adp}$ takes the rendered images $I_0$ and $I_N$ as input and generates $I_n$ where $\{ n \in \mathbb{Z} \mid 1 \leq n \leq N-1 \}$. With guidance frames $I_m$ where $\{ m \in \mathbb{Z}\mid M_1 \leq m \leq M_2 \},\; M_2 <N$, we iteratively use these frames $I_m$ during the backward denoising process. For example, at timestep $t$ during denoising, noisy latents $z_{n_t}$ are replaced with the encoded guidance frames combined with noise $\epsilon_t$, which is equivalent to producing them through the forward diffusion process. Please refer to Figure ~\ref{fig:inpaint} for illustration that replaces corresponding parts of $z_{n_t}$ with $\mathcal{E}_{vae}(I_m) + \epsilon_t$ to produce $z'_{n_t}$. This process is carried out iteratively, regenerating $I_m$ while ensuring that the interpolated results are correctly aligned with the distribution of $D_{adp}$.

\begin{figure}[h]
\centering
\includegraphics[width=1\linewidth]{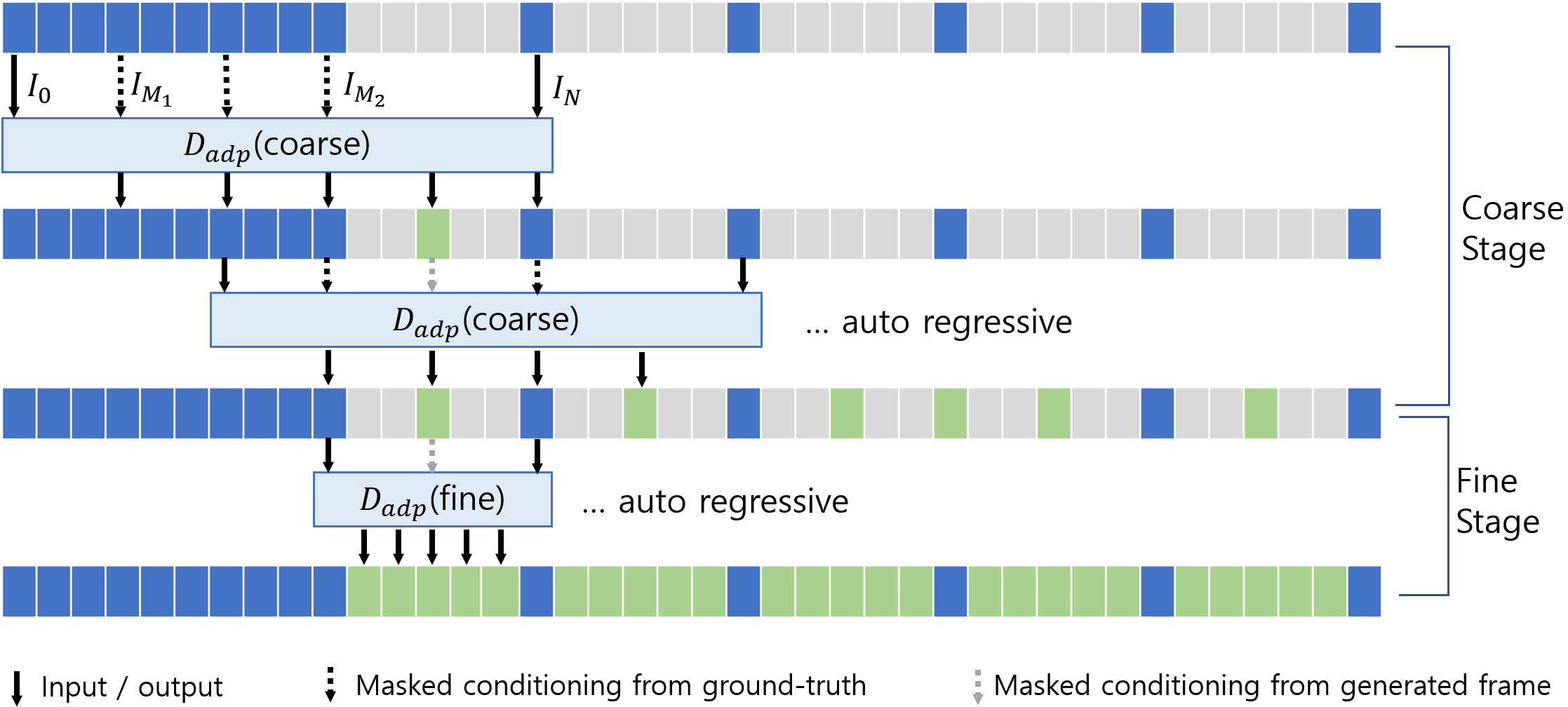}
\caption{Two stage inference of $D_{adp}$. First, at coarse stage, low frame-rate video is generated in auto regressive manner. Next, high frame-rate video is generated from low frame-rate video.}
\vspace{-2mm}
\label{fig:two_stage}
\end{figure}

After the iterative generation, the interpolated sparse video is generated. Because this video is generated with a low frame-rate with large semantic jumps across neighboring frames, the resulting motion is unlikely to be smooth. To address this, we perform the second stage for fine video generation. Similar to the first stage, we use keyframes as input but with a smaller time interval. As we additionally have the frames generated in the first stage, they can be effectively used as guidance for the second stage. This process is shown in Figure ~\ref{fig:two_stage}. Blue boxes represent given frames, green boxes represent generated frames, and gray boxes represent the frames that are yet to be generated.

\subsection{Motion-video Mimicking}\label{subsec:mimicking}
The purpose of motion video mimicking is to elevate the generated video into 3D motion data, guided by keyframes, to produce the final motion output. We achieve this by optimizing the root position $P$ and per-joint rotation $R$, starting from keyframes using our scene-specific 3D joint estimator $\mathcal{E}_{scene}$. We will first discuss the $\mathcal{E}_{scene}$ in Sec~\ref{subsec:pose} and the optimization process in Sec~\ref{subsec:optimization}.

\subsubsection{Scene-specific Joint Estimator}\label{subsec:pose}
\begin{figure}[t]
\centering
\includegraphics[width=1.05\linewidth]{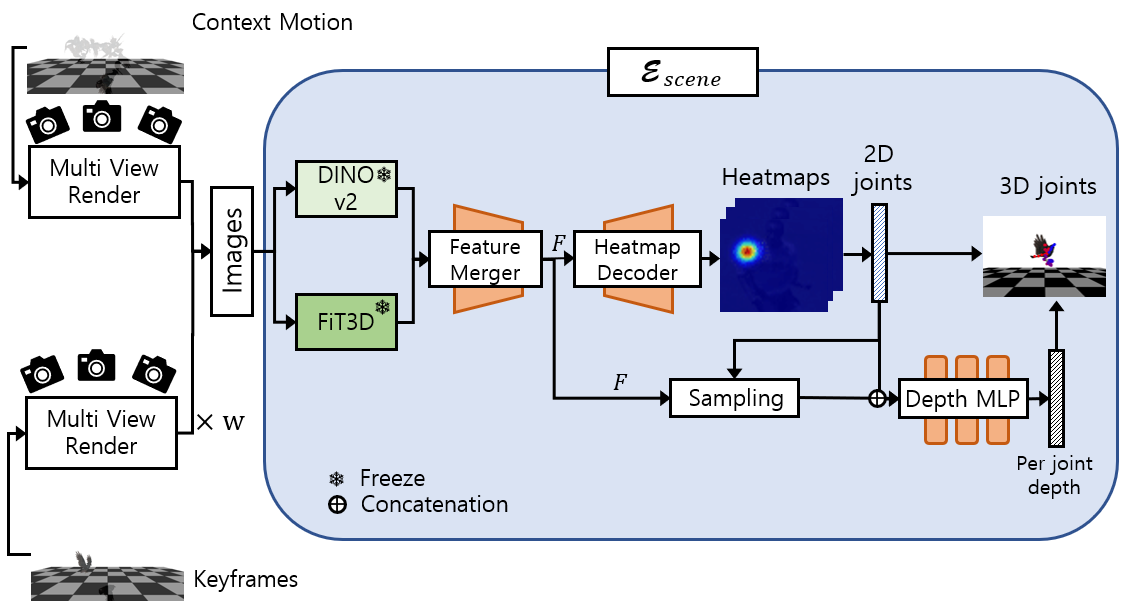}
\vspace{-2mm}
\caption{Overview of joint estimator training process.}
\vspace{-3mm}
\label{fig:joint_estimator}
\end{figure}

Accurate joint position estimation is a fundamental objective for optimizing motion in video analysis. Although there are few general-purpose 2D keypoint detection methods~\cite{yang2025x, xu2022pose}, we observed that these approaches, typically trained on real images, often fail to generalize to unseen rendered characters. To address this limitation, we propose a 3D joint estimation network $\mathcal{E}_{scene}$, specialized in a specific scene. The training process is shown in Figure~\ref{fig:joint_estimator}. We use the same dataset as in ICAdapt, supplemented with additional images for target keyframes, applying a weight $w>1$.
This weight is used to avoid overfitting to context frames, because the context frames are extracted at 30 fps, while the keyframes are at 1 fps. Moreover, we exclude images rendered from back views because only the front views will be used for motion-video mimicking. Training $\mathcal{E}_{scene}$ with back views would confuse left-right information.

Despite incorporating two seconds of context frames and keyframes, the available data remains relatively limited in size to train a 3D estimator from scratch. To bolster performance of $\mathcal{E}_{scene}$ with limited data, we leverage features from DINOv2~\cite{oquab2023dinov2}, which are trained with register tokens~\cite{darcet2023vision}. These features encapsulate rich semantic information through self-supervised training on extensive image datasets. Recognizing that DINOv2 effectively captures 2D semantic features but lacks an inherent understanding of 3D structures, we supplement DINOv2 features with FiT3D~\cite{yue2025improving} features, a 3D-aware variant that provides structural guidance. The feature merger is achieved using lightweight convolutional layers and concatenation to prevent overfitting. The resultant fused feature $F$ is subsequently decoded into heatmaps, facilitating precise 2D pose estimation.

We pass the estimated 2D joint positions through MLP layers to predict depth for each joint. Building on recent findings~\cite{zhao2024single} that emphasize the superiority of contextual features over large datasets for 3D pose estimation, we leverage $F$ also for per joint depth estimation. Specifically, we sample the feature $f_{x,y} \in \mathbb{R}^{1 \times 1 \times C}$ from $F$, which corresponds to the estimated 2D joint in spatial dimension. Then, the sampled feature $f_{x,y}$ is concatenated with the estimated 2D joint position and passed through a Depth MLP for the estimation of a per joint depth value. The training objective of $\mathcal{E}_{scene}$ is based on a MSE loss, defined as follows:
\begin{equation}
    L_{joint} = \Vert \mathcal{T}(G_{joint},{p}_{cam}) - J_{est} \Vert_2^2, \label{eq:joint}
\end{equation}
where $G_{joint}$ indicates the ground-truth 3D joint positions in global space, automatically derived using context frames and keyframes, $p_{cam}$ denotes the camera parameters, $\mathcal{T}$ is an affine transformation that projects global positions onto screen space, and $J_{est}$ indicates the estimated joint positions. Given that the depth is represented in normalized device coordinates (NDC) space, we denormalize it using the half of height to match the range of width-height and depth information. 

\subsubsection{Optimization Process}~\label{subsec:optimization}
We initiate the optimization process using two keyframes, denoted as ${P_{k_1}, R_{k_1}}$ and ${P_{k_2}, R_{k_2}}$, which represent the root positions and per-joint rotations at frames $k_1$ and $k_2$, respectively. From these keyframes, we iteratively optimize the motion parameters for interpolated frames by moving inwards toward the center. Specifically, we first optimize the parameters for the frames adjacent to the keyframes: ${P_{k_1+1}, R_{k_1+1}}$ and ${P_{k_2-1}, R_{k_2-1}}$ and for each subsequent frame, we initialize the optimization using the parameters from the nearest previously optimized frames. This optimization process continues as long as $k_1 + f < k_2 - f$, where $f$ is the number of frames optimized from each keyframe. This sequential optimization ensures fast convergence and maintains temporal coherence with adjacent frames. Because the keyframes are the only ground truth data available, they serve as the initial poses during the optimization of the surrounding frames.

In addition to 3D joint positions estimated using $\mathcal{E}_{scene}$, we utilize an image loss to minimize the image-wise difference between the rendered character and the generated video. This image loss captures subtle movements that the joint loss might miss. We also utilize two regularization losses, $L_{reg}$. Position regularization enforces the root position to be similar to the position of two nearest keyframes, while rotation regularization enforces the current rotation to be similar to that of the previous frame to stabilize the optimization process. To this end, the objective for Motion-Video Mimicking is defined as follows:
\hspace{-2mm}
\scalebox{0.85}{
\begin{minipage}{\linewidth} 
    \begin{align}
        \underset{P,R}{\text{argmin}}&\Vert \mathcal{T}(\mathcal{M}(P,R),{p}_{cam}) - \mathcal{E}_{scene}(\hat{I})\Vert_2^2  + \lambda_{img}\Vert I_{P,R} - \hat{I}\Vert_2^2  + L_{reg} \nonumber \\
        \text{where}& \;L_{reg} = \lambda_{pos}\Vert P_j - P_{intp }\Vert_2^2 + \lambda_{rot}\Vert R_j - R_{prev}\Vert_2^2.\label{eq:optimize}
    \end{align}
\end{minipage}
}\vspace{1mm}
Here, $\hat{I}$ represents a single frame from the generated video, $\mathcal{M}(P, R)$ represents the 3D positions of joints in the global coordinate system given the local rotation and global position, $p_{cam}$ and $\mathcal{T}$ follows the notations of Eq.~\ref{eq:joint}. The terms $\lambda_{img}$, $\lambda_{pos}$, and $\lambda_{rot}$ denote the weights, and $I_{P,R}$ represents the rendered image given the root position and rotation. Through this optimization process, we can obtain the final 3D positions and rotations of the target character, resembling the generated 2D video.


\begin{figure*}[h]
\centering
\vspace{-4mm}
\includegraphics[width=\linewidth]{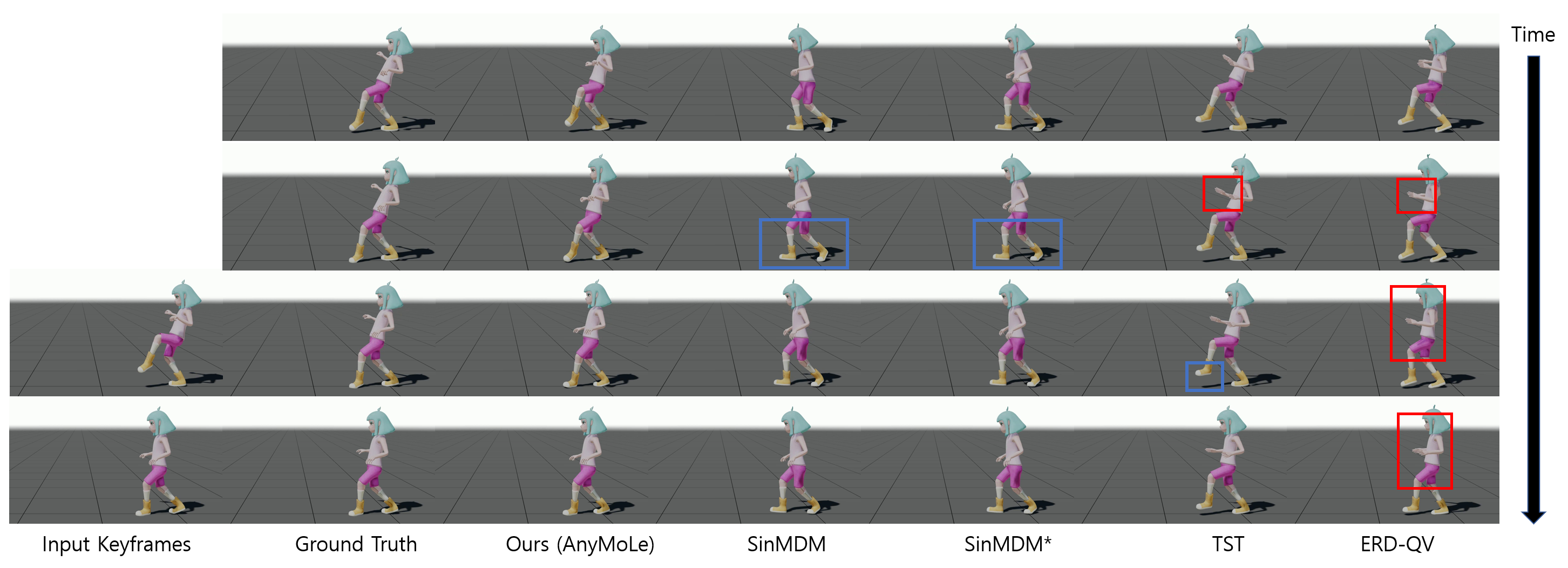}
\vspace{-6mm}
\caption{Results of baseline comparison. Our method generated in-between frames similar to the ground truth, while SinMDM and SinMDM* generated out-of-context motion (blue box). TST generated partially out-of-context footsteps (blue box) and out-of-style motion (red box). ERD-QV generated motion that has a different style, with a stiff back and sharp hand positions (red box). }
\label{fig:comparison}
\end{figure*}

\begin{table*}
\centering
\renewcommand{\arraystretch}{1.05} 
\setlength{\tabcolsep}{3pt}
\caption{Quantitative results compared with the baselines. Ours outperformed all competitors with by margin.}\vspace{-3.5mm}
\small
\begin{tabular}{|l|ccccccc|ccccccc|}
\noalign{\smallskip}\noalign{\smallskip}\hline
Character & \multicolumn{7}{c|}{Humanoid} & \multicolumn{7}{c|}{Non-humanoid}\\ \hline
Methods & HL2Q$\downarrow$ & L2Q$\downarrow$ & L2P$\downarrow$ & NPSS$\downarrow$ & LPIPS$\downarrow$ & CLIP$\uparrow$ & SSIM$\uparrow$ & HL2Q$\downarrow$ & L2Q$\downarrow$ & L2P$\downarrow$ & NPSS$\downarrow$ & LPIPS$\downarrow$ & CLIP$\uparrow$ & SSIM$\uparrow$ \\\hline
Ours & \textbf{0.0015} & \textbf{0.0011} & \textbf{0.0063} & \textbf{0.0726} & \textbf{0.0547} & \textbf{0.9711} & \textbf{0.9343} & \textbf{0.0019} & \textbf{0.0021} & \textbf{0.0299} & \textbf{0.1863} & \textbf{0.0433} & \textbf{0.9675} & \textbf{0.9531} \\ \hline
SinMDM & 0.0971 & 0.0386 & 0.3733 & 1.7148 & 0.1007 & 0.9393 & 0.9185 & 0.2465 & 0.1710 & 0.1382 & 4.4324 & 0.0816 & 0.9536 & 0.9362 \\
SinMDM* & 0.0981 & 0.0390 & 0.3991 & 1.7584 & 0.1053 & 0.9287 & 0.9161 & 0.2467 & 0.1713 & 0.1973 & 4.4076 & 0.0943 & 0.9495 & 0.9332 \\
TST & 0.0028 & 0.0106 & 0.0209 & 4.3509 & 0.0953 & 0.9606 & 0.9148 & - & - & - & - & - & - & - \\
ERD-QV & 0.0028 & 0.0109 & 0.0098 & 4.3801 & 0.0693 & 0.9656 & 0.9272 & - & - & - & - & - & - & - \\
\hline
\end{tabular}
\vspace{-2mm}
\label{Tab:qualitative}
\end{table*}

\section{Experiments}
\subsection{Implementation Details}
We implemented AnyMoLe and conducted all training and inference on a computer with an Nvidia A6000 GPU. For ICAdapt, frames are rendered in four different views ($N=4$; front, left, right, and back), while the back view is discarded for $\mathcal{E}_{scene}$. ICAdapt was conducted for 500 steps with a batch size of 16, while $\mathcal{E}_{scene}$ was trained for 3,500 steps with a batch size of 32. Weight $w$ for keyframes in Figure~\ref{fig:joint_estimator} was set to 3. For two-stage inference, the first stage generated a 5fps video, while the second stage generated a 15fps video. After optimization on the 15fps video, we upsampled the video to 30fps for evaluation by applying a Gaussian filter and slerp. Weights $\lambda_{img}$, $\lambda_{pos}$, and $\lambda_{rot}$ in Eq.\ref{eq:optimize} were set to 50, 7,000, and 30,000, respectively. For all of our experiments, we used the first two seconds of the input motion as the context motion and sampled the remaining portion of the motion at one-second intervals for the target keyframes.

\subsection{Evaluation Metrics}
For evaluation metrics, we measured the rotation, position, and rendered image similarity, all compared with ground-truth motion. For rotation, we used L2Q and NPSS metrics. L2Q is the L2 distance in local quaternion space compared to ground truth, while NPSS is the angular frequency similarity proposed in~\citet{gopalakrishnan2019neural}. Because the joints that are near the root have a larger impact on overall character motion compared to leaf joints, we also propose to use an adjunctive metric, Hierarchy-filtered L2Q (HL2Q). We conducted the filtering process using the known skeletal hierarchy. In our experiments, we used the child joints only up to 50\% of the depth from the root joint in the skeleton hierarchy. Additional results with varying filtering thresholds are presented in the supplementary material. For positional differences, we used L2P, which measures the L2 distance between the global joint position and ground truth. Lastly, we used LPIPS~\cite{zhang2018unreasonable} for perceptual similarity, CLIP~\cite{radford2021learning} for semantic similarity, and SSIM for structural similarity of rendered characters, compared with the ground truth under the same camera setting.

\begin{figure*}[h]
\centering
\vspace{-3mm}
\includegraphics[width=0.9\linewidth]{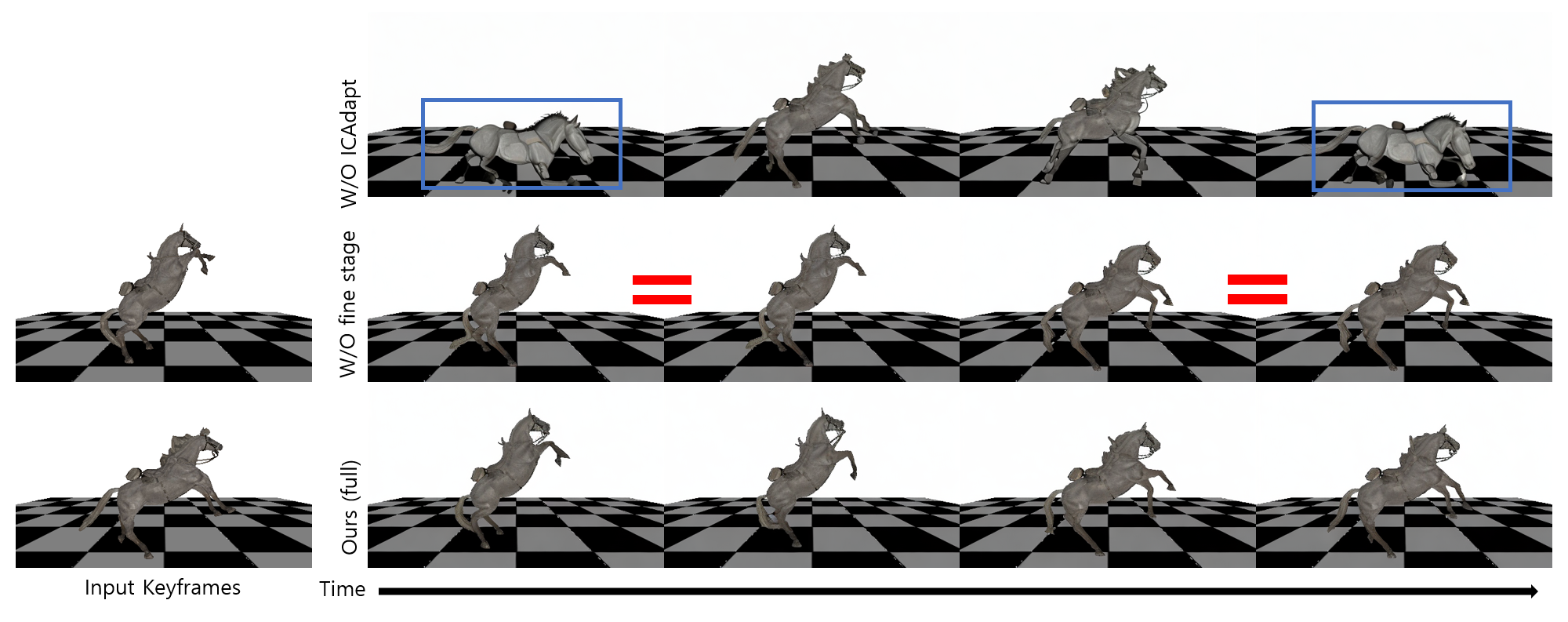}
\vspace{-2mm}
\caption{Ablation results on video generation. Without applying ICAdapt, each frame of the video exhibited inconsistencies, such as generating noticeable style shifts (blue box). Omitting the fine-stage process resulted in a low frame rate, making identical or significant jumps between frames.}
\vspace{-2mm}
\label{fig:ablation_add}
\end{figure*}

\subsection{Comparison with Baselines}
We conducted comparisons with three different baselines: ERD-QV~\cite{harvey2020robust}, TST~\cite{qin2022motion}, and SinMDM~\cite{zhao2024single}. ERD-QV and TST are motion in-betweening methods trained on large MOCAP datasets, while SinMDM is a motion generation method trained on single motions, which demonstrated motion in-betweening as its application. For ERD-QV and TST, their original MOCAP datasets~\cite{harvey2020robust} were used, while SinMDM was trained using the same context frames as our method. We also made a variant of SinMDM, SinMDM* that preserves keyframes by omitting region-of-interest (ROI) blending. This adjustment was made because ROI blending was originally designed for settings involving multiple target frames.

In our experiments, we used 20 different motions of diverse characters, including both humanoid characters and non-humanoid characters (e.g., birds, snakes, and dinosaurs). The humanoid characters were sourced from Mixamo~\cite{mixamo}, while the non-humanoid characters were obtained from Truebones Zoo~\cite{truebones2024}. Because the ERD-QV and TST methods were trained for a single character, we retargeted their motions to target humanoid characters using the commercial software MotionBuilder~\cite{motionbuilder}. 
However, as these methods cannot generate motions for non-humanoid characters such as snakes, comparison for non-humanoid characters was not performed.

Comparison results are shown in Figure~\ref{fig:comparison}. While our method faithfully generated frames following the style of the input keyframes, SinMDM and SinMDM* resulted in motion that was out of context due to the method overfitting to the given context frames. This is because the original SinMDM trains the model with a full motion; the context frames may not contain all the motions that will be used for the motion in-betweening task. For ERD-QV and TST, while these models generated the foot motion better than SinMDM, they failed to follow the style of the given motion, such as upper body tilt or hand positions. Quantitative results are presented in Table~\ref{Tab:qualitative}.  In all metrics, our method outperformed competitors by large margin.


\subsection{Ablation Study}

\paragraph{Video Generation}
The first stage of AnyMoLe is to generate a video by filling in the motion in 2D using the knowledge of a large video diffusion model. In this paper, we proposed ICAdapt to bridge the gap between real-world videos and rendered scenes. Additionally, we implemented a two-stage generation process to achieve a higher frame rate, ensuring smooth motion optimization, as frame-wise similarity aids in optimizing between neighboring frames. We conducted an ablation study without these components. As shown in Figure~\ref{fig:ablation_add}, when ICAdapt was not used, spatial information—such as the style of horse-changed (blue box). When the fine-stage was not used, a low frame rate video was produced, with identical nearby frames. In contrast, our method produced results with smooth transitions and no visual artifacts. 

To verify whether degraded video quality indeed negatively impacts the generated motion, we conducted a quantitative evaluation on the full process without these components. For this quantitative ablation study, we used the same dataset and metrics as in the baseline comparison, but this time the resulting values for humanoid and non-humanoid characters were averaged for simplification. As shown in the results for \textit{w/o} ICAdapt and \textit{w/o} fine stage in Table~\ref{Tab:ablation}, the final motion extracted from lower-quality videos led to lower performance on all metrics. This occurred because the degraded video quality negatively affects the optimization process by making joint estimation more challenging for \textit{w/o} ICAdapt and causing larger motion gaps between adjacent frames for \textit{w/o} fine stage.

\begin{table}
\hspace{-2mm}
\vspace{-4mm}
\renewcommand{\arraystretch}{1.3} 
\setlength{\tabcolsep}{2pt}
\caption{Quantitative results of ablation study.}
\footnotesize
\vspace{-1mm}
\begin{tabular}{|l|ccccccc|}
\noalign{\smallskip}\noalign{\smallskip}\hline
Methods & HL2Q$\downarrow$ & L2Q$\downarrow$ & L2P$\downarrow$ & NPSS$\downarrow$ & LPIPS$\downarrow$ & CLIP$\uparrow$ & SSIM$\uparrow$ \\\hline
Ours & \textbf{0.00169} & \textbf{0.00158} & \textbf{0.01811} & 0.1295 & \textbf{0.04901} & \textbf{0.9693} & \textbf{0.9437} \\
\textit{w/o} ICAdapt & 0.00210 & 0.00181 & 0.02640 & 0.1413 & 0.07115 & 0.9619 & 0.9340 \\
\textit{w/o} fine stage & 0.00200 & 0.00171 & 0.02204 & 0.1345 & 0.06037 & 0.9653 & 0.9385 \\
\textit{w} XPose & 0.00174 & 0.00167 & 0.01976 & \textbf{0.1288} & 0.06004 & 0.9678 & 0.9421 \\
\textit{w/o} data select & 0.00799 & 0.00421 & 0.15460 & 0.2434 & 0.12735 & 0.9061 & 0.9211 \\
\hline
\end{tabular}
\vspace{-2mm}
\label{Tab:ablation}
\end{table}

\paragraph{Motion Video Mimicking}
For Motion Video Mimicking, we first trained our scene-specific joint estimator and sequentially optimized 3D joint positions to estimate the character motion. We compared our joint estimator with general pose esitimation method XPose~\cite{yang2025x}. As shown in Table~\ref{Tab:ablation}, ours outperformed in all the metrics except NPSS. This is because XPose, trained with real images, sometimes failed to estimate the positions correctly when the appearance of the character is somewhat stylized. We additionally conducted a study on our data selection process, which differs from the training dataset used for ICAdapt. Specifically, we included rendered keyframes with weight $w$ but excluded back-view images when training the scene-specific joint estimator. This data choice was crucial because excluding keyframes led to worse results due to overfitting to context frames as indicated by the last row in Table~\ref{Tab:ablation}. Additionally, because motion video mimicking was performed only in the frontal view, training with an in-domain dataset helped improve performance.

\subsection{User Study}
We conducted a user study with 21 participants, 11 female and 10 male
aged 27.1 on average to evaluate motion in-betweening results based on human perception. Each participant was presented with three questions on each of 20 motions and asked to choose the one that is: 1) most similar to the ground-truth motion (Similar), 2) best follows the last target keyframe (Faithful), and 3) most smooth and natural (Natural). The scores reported in Table~\ref{Tab:user}
represent the percentage of selections for each category. Our method received higher scores compared to the baseline methods for both humanoid and non-humanoid characters.

\begin{table}[t]
\centering
\renewcommand{\arraystretch}{0.95}
\setlength{\tabcolsep}{1mm}
\caption{Perceptual study results on ground-truth similarity, target keyframe faithfulness, and motion naturalness. The number indicates the percentage of selections.}\small \vspace{-2mm}
\begin{tabular}{|l|ccc|ccc|}
\hline
Character & \multicolumn{3}{c|}{Humanoid} & \multicolumn{3}{c|}{Non-humanoid}\\ \hline
Methods  & Similar & Faithful & Natural & Similar & Faithful & Natural  \\
\hline
Ours     & 60.12 & 63.10 & 64.88 & 90.48 & 92.46 & 91.67 \\ 
SinMDM      & 13.10 & 14.29  & 7.74 & 3.97  & 3.17  & 3.57  \\
SinMDM*       & 5.36  & 4.17  & 5.95  & 5.56  & 4.37  & 4.76  \\
TST  & 16.07 & 12.50 & 16.67 & -     & -     & -     \\
ERD-QV        & 5.36  & 5.95  & 4.76  & -     & -     & -     \\ 
\hline
\end{tabular}
\label{Tab:user}
\end{table}

\section{Applications}
\begin{figure}[t]
\centering
\vspace{-3mm}
\includegraphics[width=0.9\linewidth]{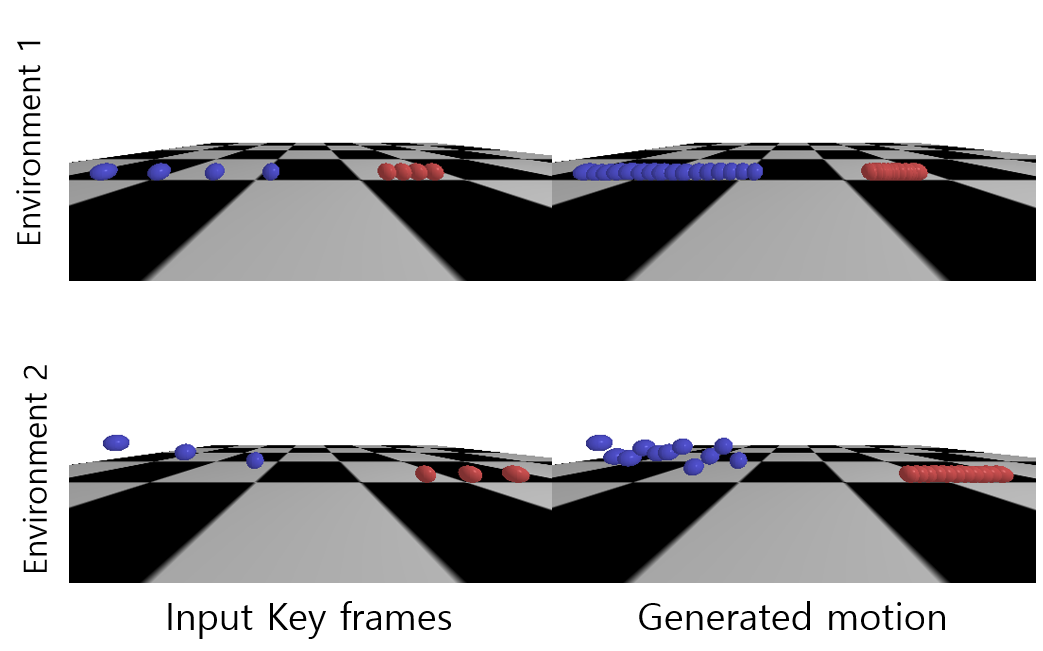}
\vspace{-3mm}
\caption{From understanding of the context, AnyMoLe can naturally generate in-between frames in a multi-object scenario.}
\label{fig:application}
\end{figure}

AnyMoLe can extend its capabilities from single character motion in-betweening to handling multi-object scenarios, such as two ball simulation. By leveraging video diffusion models with contextual understanding, it ensures smooth transitions between objects while maintaining coherent spatial relationships. This is achieved with simple modification to $\mathcal{E}_{scene}$ and optimization process, to estimate and optimize all positions of objects instead of the joints. As shown in Figure~\ref{fig:application}, AnyMoLe smoothly generates two ball simulation with different bounciness.

\section{Discussion and Conclusion}

\begin{figure}[t]
\centering
\includegraphics[width=\linewidth]{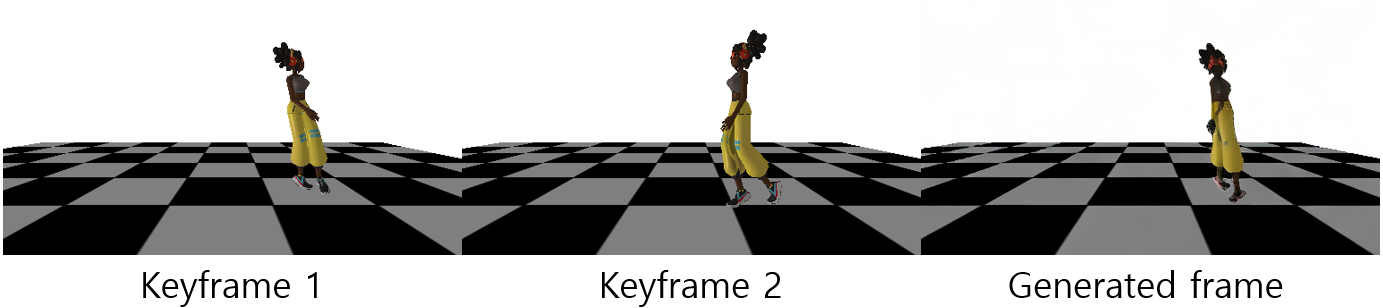}
\vspace{-5mm}
\caption{Visualization of a generated frame from a fast turn-around, showing ambiguity that can hinder joint estimation.}
\label{fig:limiation}
\vspace{-5mm}
\end{figure}

We proposed a novel motion in-betweening method to overcome the limitations posed by requirements of datasets. While our method requires training only a lightweight pose estimator using pretrained feature extractors and finetuning the video diffusion model on a small dataset, the overall process still necessitates five to six hours to complete the entire process. Additionally, for motions with fast or complex dynamics as shown in Figure~\ref{fig:limiation}, rapid movements can blur the generated videos, causing ambiguity—especially in distinguishing left from right joints—and thus hampering motion estimation. One possible direction to mitigate time requirements and improve robustness against a few ambiguous frames in generated videos is to train a context-informed, character-agnostic 3D joint estimator. If such pretraining can be achieved without sacrificing performance, it would significantly reduce computational time for training Scene-specific joint estimator and ensure robustness, even when encountering blurry frames, due to its contextual understanding.

In this paper, we introduced a novel method for motion in-betweening that effectively leverages video diffusion models to overcome the limitation of requiring character-specific dataset, which has been overlooked. To acheive this, we first identified three primary challenges associated with the naive application of these video diffusion models to motion in-betweening tasks: 1) limited contextual understanding, 2) domain gaps, and 3) difficulties in tracking motion from generated videos. To address these challenges, we introduced a two-stage video generation process using context frames, the ICAdapt method, motion-video mimicking, and a scene-specific joint estimator. By addressing these key challenges, our approach is the first to successfully achieve motion in-betweening for arbitrary characters without the need for external data. We believe that our contributions broaden the use of video generation models and will stimulate further research in 3D character motion synthesis, particularly for characters that are challenging to capture with MOCAP or to animate manually.

\clearpage

\section*{Acknowledgements}
This work was supported by the National Research Foundation of Korea (NRF) grant funded by the Korea government (MSIT) (RS-2024-00333478).

{
    \small
    \bibliographystyle{ieeenat_fullname}
    \bibliography{main}
}

\end{document}